\documentclass[12pt, letterpaper, onecolumn, peerreview]{IEEEtran}
\usepackage{exscale}
\usepackage{color}
\usepackage{colordvi}
\usepackage{graphics}
\usepackage{amsmath}
\usepackage{amssymb}
\usepackage{epsfig}
\usepackage{epic}

\newtheorem{theorem}{Theorem}[section]

\newtheorem{definition}[theorem]{Definition}
\newtheorem{corollary}{Corollary}[section]

\begin{document}
\title{How to beat the sphere-packing bound with feedback} 

\author{\authorblockN{Anant Sahai}
\authorblockA{Wireless Foundations, Dept.~of EECS \\
University of California at Berkeley \\
Email: sahai@eecs.berkeley.edu}}
\maketitle

\begin{abstract} 
The sphere-packing bound $E_{sp}(R)$ bounds the reliability function
for fixed-length block-codes. For symmetric channels, it remains a
valid bound even when strictly causal noiseless feedback is allowed
from the decoder to the encoder. To beat the bound, the problem must
be changed. While it has long been known that variable-length block
codes can do better when trading-off error probability with expected
block-length, this correspondence shows that the {\em fixed-delay}
setting also presents such an opportunity for generic channels.

While $E_{sp}(R)$ continues to bound the tradeoff between bit error
and fixed end-to-end latency for symmetric channels used {\em without}
feedback, a new bound called the ``focusing bound'' gives the limits
on what can be done with feedback. If low-rate reliable flow-control
is free (ie. the noisy channel has strictly positive zero-error
capacity), then the focusing bound can be asymptotically achieved.
Even when the channel has no zero-error capacity, it is possible to
substantially beat the sphere-packing bound by synthesizing an
appropriately reliable channel to carry the flow-control information.
\end{abstract}

\IEEEpeerreviewmaketitle

\section{Introduction}
The two most fundamental parameters when it comes to reliable data
transport are end-to-end system delay and the probability of
error. Error probability is fundamental because a low probability of
bit error lies at the heart of the digital revolution justified by the
source/channel separation theorem. Delay is important because it is
the most basic cost that a system must pay in exchange for reliability
--- it allows the laws of large numbers to be harnessed to smooth out
the variability introduced by random communication channels.

Traditionally, block-length has been used as a proxy for end-to-end
delay since block-codes are easier to understand than non-block codes.
Even when fixed end-to-end delay is desired, this paper shows that
nonblock codes can provide a tremendous advantage when feedback is
allowed. This short correspondence is a companion to our longer work
in \cite{OurUpperBoundPaper}. Some key results are reviewed here in
the next section, but the reader is referred to
\cite{OurUpperBoundPaper} for more details, motivation, as well as a
perspective on the existing results in the literature. The new
contribution in this correspondence comes in
Section~\ref{sec:syntheticflow}. It shows how to construct a special
fixed-delay code over a DMC using noiseless feedback. It beats the
sphere-packing bound with fixed-delay in the high-rate regime even for
channels (like the BSC) that have no zero-error capacity. A plot is
given for the BSC-0.4 case that provides an explicit counterexample to
Pinsker's assertion (Theorem~8 in \cite{PinskerNoFeedback}) that this
is impossible to do.

Simsek had earlier built codes for the BSC in \cite{TuncThesis,
  SimsekJainVaraiya} which beat the sphere-packing bound with fixed
delay. Those were fundamentally built upon the equivalence between
scalar stabilization problems and feedback communication problems
established in \cite{SahaiThesis, ControlPartI}, but were hard to
analyze. They also did not work at high rates. The advantage of the
codes given here is their conceptual simplicity and the fact that they
beat the sphere-packing bound in the high-rate regime. These codes do
not do well in the low-rate regime and it is clear that they could be
married with Simsek-codes to give some improvements at low-rate.
However, this would not be enough to reach the focusing bound so there
is much room for improvement.

\section{Review} \label{sec:background}

\subsection{Block coding}
The fundamental lower-bound on error probability comes from the
sphere-packing or volume bound, and this bound is also known to be
achievable at high rates by random-coding \cite{gallager}. Reliable
communication is not possible if during the block, the channel acts
like one whose capacity is less than the target rate. Following
\cite{csiszarkorner} and \cite{Haroutunian}, for block codes this idea
immediately gives the following bound on the exponential error
probability:
\begin{equation} \label{eqn:primitiveupperbound}
E^+(R) = \inf_{G: C(G) < R} 
            \max_{\vec{r}} D\left(G||P | \vec{r}\right)
\end{equation}
where $D(G || P | \vec{r})$ is the divergence term that governs the
exponentially small probability of the true channel $P$ behaving like
channel $G$ when facing the input distribution coming from the
codeword composition $\vec{r}$. 

Even with causal noiseless feedback, there is no way around this bound
because channel capacity does not increase with feedback for
memoryless channels. Without feedback, the bound can be tightened to
the form traditionally known as the sphere-packing bound.
\begin{equation} \label{eqn:spherepackingbound}
    E_{sp}(R) = \max_{\vec{r}} \min_{G: I(\vec{r},G) \leq R } D\left(G||P|\vec{r}\right)
\end{equation}
For symmetric channels, the optimizing codeword composition $\vec{r}$
is always uniform and $E_{sp}(R) = E^+(R)$. Thus, for fixed-block codes and
symmetric DMCs, no only does causal feedback not improve capacity, but
it does not improve reliability either\cite{DobrushinReliability}!

An alternate form for $E_{sp}(R)$ is given by:
\begin{equation} \label{eqn:rhospherepack}
  E_{sp}(R) = \max_{\rho > 0} \big[ E_0(\rho) - \rho R \big]
\end{equation}
with the Gallager function $E_0(\rho)$ defined as:
\begin{equation} \label{eqn:enought}
  E_0(\rho) = \max_{\vec{q}} -\ln \sum_y \bigg[ \sum_x q_x
  p_{x,y}^\frac{1}{1+\rho} \bigg]^{(1+\rho)}
\end{equation}
Note that for symmetric channels, it suffices to use a uniform
$\vec{q}$ while optimizing (\ref{eqn:enought}). Also, since the
random-coding error exponent is given by:
\begin{equation} \label{eqn:rhorandom}
  E_{r}(R) = \max_{0 < \rho \leq 1} \big[ E_0(\rho) - \rho R \big]
\end{equation}
It is clear that the sphere-packing bound is achievable, even without
feedback, at rates close to $C$ since for those rates, $\rho < 1$
optimizes both expressions \cite{gallager}. The points on the
sphere-packing bound where $\rho > 1$ are also achievable by random
coding if the sense of ``correct decoding'' is slightly
relaxed. Rather than forcing the decoder to emit a single estimated
codeword, list-decoding allows the decoder to emit a list of guessed
codewords. The decoding is considered correct if the true codeword is
on the list. For list-decoding with list size $\ell$ in the context of
random codes, Problem 5.20 in \cite{gallager} reveals that
\begin{equation} \label{eqn:listrhorandom}
  E_{r,\ell}(R) = \max_{0 < \rho \leq \ell} \big[ E_0(\rho) - \rho R \big]
\end{equation}
is achievable. At high rates (where the maximizing $\rho$ is small),
there is no benefit from relaxing to list-decoding, but it makes a
difference at low rates.

\subsection{Non-block codes} \label{sec:intrononblock}

Another classical approach to the problem of reliable communication is
to consider codes without a block structure. Convolutional and tree
codes represent the prototypical examples. It was realized early on
that in an infinite\footnote{More precisely, these are unbounded
constraint length codes since at any finite time there are only a
finite number of data bits so far.} constraint length convolutional
code under ML decoding, all bits will eventually be decoded correctly
\cite{gallager}. However, if the end-to-end delay is forced to be
bounded, then the bit error probability with delay is governed by
$E_{r}(R)$ for random convolutional codes, even when the constraint
lengths are unbounded \cite{ForneyML}. This performance with delay is
also achievable using an appropriately biased sequential decoder
\cite{JelinekSequential}. A nice feature of sequential decoders is
that they are not tuned to any target delay --- they can be prompted
for estimates at any time and they will give the best estimate that
they have. Thus an infinite constraint-length convolutional code with
appropriate sequential decoding achieves the exponent $E_r(R)$ delay
universally over all (sufficiently long) delays.

Pinsker claimed in \cite{PinskerNoFeedback} that the sphere-packing
bound continued to bound the performance of nonblock codes both with
and without feedback. He had proofs for the BSC case, but asserted
that the result held more generally. While he was right for the
without feedback case, it turns out that there is a subtle flaw in his
argument regarding the case with feedback. 

\subsubsection{The BEC example} \label{sec:becexample}

This example, repeated from \cite{OurUpperBoundPaper} for the
reviewer's convenience, shows the power of feedback in the delay
context. The binary erasure channel with erasure probability $\delta <
\frac{1}{2}$ used at bit-rate $R' = \frac{1}{2}$ gives a
counterexample to Pinsker's conjecture. The BEC is so simple that
everything can be understood with a minimum of overhead.

\begin{equation} \label{eqn:becspherebound}
 E_{sp}(\frac{1}{2}) = D(\frac{1}{2}||\delta) = 
                     - \frac{\ln(4\delta(1-\delta))}{2}
\end{equation}
For $\delta = 0.4$, this corresponds to an error exponent of about
$0.02$. Even with feedback, there is no way for a fixed block-length
code to beat this exponent. If the channel lets fewer than
$\frac{n}{2}$ bits through the channel, it is impossible to reliably
communicate an $\frac{n}{2}$ bit message!

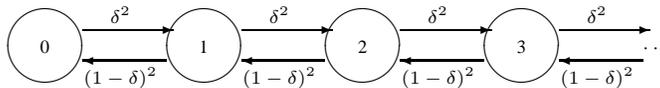
\begin{figure}
\begin{center}
\setlength{\unitlength}{2500sp}%
\begingroup\makeatletter\ifx\SetFigFont\undefined%
\gdef\SetFigFont#1#2#3#4#5{%
  \reset@font\fontsize{#1}{#2pt}%
  \fontfamily{#3}\fontseries{#4}\fontshape{#5}%
  \selectfont}%
\fi\endgroup%
\begin{picture}(6458,816)(518,-494)
{\color[rgb]{0,0,0}\thinlines
\put(5626,-61){\circle{750}}
}%
{\color[rgb]{0,0,0}\put(4051,-61){\circle{750}}
}%
{\color[rgb]{0,0,0}\put(2476,-61){\circle{750}}
}%
{\color[rgb]{0,0,0}\put(901,-61){\circle{750}}
}%
{\color[rgb]{0,0,0}\put(6001, 89){\vector( 1, 0){900}}
}%
{\color[rgb]{0,0,0}\put(6826,-211){\vector(-1, 0){825}}
}%
{\color[rgb]{0,0,0}\put(4426, 89){\vector( 1, 0){900}}
}%
{\color[rgb]{0,0,0}\put(5251,-211){\vector(-1, 0){825}}
}%
{\color[rgb]{0,0,0}\put(2851, 89){\vector( 1, 0){900}}
}%
{\color[rgb]{0,0,0}\put(3676,-211){\vector(-1, 0){825}}
}%
{\color[rgb]{0,0,0}\put(1276, 89){\vector( 1, 0){900}}
}%
{\color[rgb]{0,0,0}\put(2101,-211){\vector(-1, 0){825}}
}%
\put(6376,164){\makebox(0,0)[b]{\smash{\SetFigFont{7}{6}{\rmdefault}{\mddefault}{\updefault}{\color[rgb]{0,0,0}$\delta^2$}%
}}}
\put(6376,-436){\makebox(0,0)[b]{\smash{\SetFigFont{7}{6}{\rmdefault}{\mddefault}{\updefault}{\color[rgb]{0,0,0}$(1-\delta)^2$}%
}}}
\put(4801,164){\makebox(0,0)[b]{\smash{\SetFigFont{7}{6}{\rmdefault}{\mddefault}{\updefault}{\color[rgb]{0,0,0}$\delta^2$}%
}}}
\put(4801,-436){\makebox(0,0)[b]{\smash{\SetFigFont{7}{6}{\rmdefault}{\mddefault}{\updefault}{\color[rgb]{0,0,0}$(1-\delta)^2$}%
}}}
\put(3226,164){\makebox(0,0)[b]{\smash{\SetFigFont{7}{6}{\rmdefault}{\mddefault}{\updefault}{\color[rgb]{0,0,0}$\delta^2$}%
}}}
\put(3226,-436){\makebox(0,0)[b]{\smash{\SetFigFont{7}{6}{\rmdefault}{\mddefault}{\updefault}{\color[rgb]{0,0,0}$(1-\delta)^2$}%
}}}
\put(1651,164){\makebox(0,0)[b]{\smash{\SetFigFont{7}{6}{\rmdefault}{\mddefault}{\updefault}{\color[rgb]{0,0,0}$\delta^2$}%
}}}
\put(1651,-436){\makebox(0,0)[b]{\smash{\SetFigFont{7}{6}{\rmdefault}{\mddefault}{\updefault}{\color[rgb]{0,0,0}$(1-\delta)^2$}%
}}}
\put(2476,-136){\makebox(0,0)[b]{\smash{\SetFigFont{7}{6}{\rmdefault}{\mddefault}{\updefault}{\color[rgb]{0,0,0}1}%
}}}
\put(901,-136){\makebox(0,0)[b]{\smash{\SetFigFont{7}{6}{\rmdefault}{\mddefault}{\updefault}{\color[rgb]{0,0,0}0}%
}}}
\put(4051,-136){\makebox(0,0)[b]{\smash{\SetFigFont{7}{6}{\rmdefault}{\mddefault}{\updefault}{\color[rgb]{0,0,0}2}%
}}}
\put(5626,-136){\makebox(0,0)[b]{\smash{\SetFigFont{7}{6}{\rmdefault}{\mddefault}{\updefault}{\color[rgb]{0,0,0}3}%
}}}
\put(6976,-136){\makebox(0,0)[b]{\smash{\SetFigFont{7}{6}{\rmdefault}{\mddefault}{\updefault}{\color[rgb]{0,0,0}$\cdots$}%
}}}
\end{picture}
\end{center}
\caption{The birth-death Markov chain governing the rate $\frac{1}{2}$
feedback communication system over an erasure channel.}
\label{fig:birthdeath}
\end{figure}

If causal noiseless feedback is available, the natural nonblock code
just retransmits a bit until it is correctly received. As bits arrive
steadily at the rate $R' = \frac{1}{2}$, they enter a FIFO queue of
bits awaiting transmission. If we look at the queue state every two
channel uses, it can be modeled (see Figure~\ref{fig:birthdeath}) as a
birth-death Markov chain with a $\delta^2$ probability of birth and a
$(1-\delta)^2$ probability of death. Converting that into an error
exponent with delay $d$ gives: 
\begin{equation} \label{eqn:halfbecfeedback}
 E^{bec}_{f}(\frac{1}{2}) = \ln(1-\delta) - \ln(\delta)
\end{equation}
Plugging in $\delta=0.4$ gives an exponent of more than $0.40$. This
is about twenty times higher than the sphere-packing bound! 

\subsection{The focusing bound}

Restricting attention to symmetric channels, the BEC case can be
abstracted to get a general bound on the probability of error with
delay. \cite{OurUpperBoundPaper} calls this bound the ``focusing
bound'' because it is based on the idea of having the encoder focus as
much of the decoder's uncertainty as possible onto bits whose
deadlines are not pending. 

\begin{definition} \label{def:feedbackencoder}
A {\em rate $R$ encoder with noiseless feedback} is a sequence of maps
${\cal E}_t$. The range of each map is the discrete set $\cal X$. The
$t$-th map takes as input the available data bits $B_1^{\lfloor R't
\rfloor}$, as well as all the past channel outputs $Y_1^{t-1}$. 

{\em Randomized encoders with noiseless feedback} also have access to
a continuous uniform random variable $W_t$ denoting the common
randomness available in the system.
\end{definition}
\vspace{0.1in}
\begin{definition} \label{def:decoder}
A {\em delay $d$ rate $R$ decoder} is a sequence of maps ${\cal
D}_i$. The range of each map is just an estimate $\widehat{B}_i$ for the
$i$-th bit taken from $\{0,1\}$. The $i$-th map takes as input the
available channel outputs $Y_1^{\lceil \frac{i}{R'}
\rceil + d}$ which means that it can see $d$ time units beyond when
the bit to be estimated first had a chance to impact the channel
inputs.

{\em Randomized decoders} also have access to all the continuous
uniform random variables $W_t$. 
\end{definition}
\vspace{0.1in}

\begin{definition} \label{def:achievable}
The fixed-delay error exponent $\alpha$ is asymptotically {\em
achievable} at rate $R$ across a noisy channel if for every delay
$d_j$ in some increasing sequence $d_j \rightarrow \infty$ there
exist rate $R$ encoders and delay $d_j$ decoders ${\cal E}^{d_j},
{\cal D}^{d_j}$ that satisfy the following properties when used with
input bits $B_i$ drawn from iid fair coin tosses.
\begin{enumerate} 
 \item For every $j$, there exists an $\epsilon_j < 1$ so that $P(B_i
       \neq \widehat{B}_i(d_j)) \leq \epsilon_j$ for every $i \geq 1$. The
       $\widehat{B}_i(d_j)$ represents the delay $d_j$ estimate of $B_i$
       produced by the ${\cal E}^j, {\cal D}^j$ pair connected to the
       input $B$ and the channel in question.

 \item $\lim_{j \rightarrow \infty} \frac{-\ln \epsilon_j}{d_j} \leq
       \alpha$
\end{enumerate}

The exponent $\alpha$ is asymptotically {\em achievable universally
over delay} or in an {\em anytime fashion} if a single encoder ${\cal
E}$ can be used above for all $d_j$ above.
\end{definition}
\vspace{0.1in}

\begin{theorem} {\em Focusing bound from \cite{OurUpperBoundPaper}:}
\label{thm:generalfeedbackbound} For a discrete memoryless channel,
no delay exponent $\alpha > E_{a}(R)$ is asymptotically achievable
even if the encoders are allowed access to noiseless feedback.
\begin{equation} \label{eqn:generalfeedbackbound}
E_a(R)  = \inf_{0 < \lambda < 1} \frac{E^+(\lambda R)}{1-\lambda}
\end{equation}
where $E^+$ is the Haroutunian exponent from
(\ref{eqn:primitiveupperbound}). When the DMC is symmetric, $E_a(R)$ can
be expressed parametrically as:
\begin{equation} \label{eqn:symmetricfeedbackbound}
 E_a(R) = E_0(\eta)\,\,;\,\, R = \frac{E_0(\eta)}{\eta}
\end{equation}
where $E_0(\eta)$ is the Gallager function from (\ref{eqn:enought}),
and $\eta$ ranges from $0$ to $\infty$.
\end{theorem}
\vspace{0.1in} 

\subsection{The $(n,c,l)$ family of codes}
The focusing bound is attained for the BEC with feedback using the
natural ``repeat bits until successful'' code. As demonstrated in
\cite{OurUpperBoundPaper}, it can also be asymptotically attained for
any noisy channel provided we have access to a low-rate channel that
can deliver perfectly noiseless flow-control bits from the encoder to
the decoder. The code is reviewed below.
 
Call $c \geq 1$ the chunk length, $2^l$ the list length, and $n > l$
the data block length. The $(n, c, l)$ scheme is:
\begin{itemize} 
  \item Queue up incoming bits and assemble them into blocks of size
        $\frac{n c R}{\ln 2}$ bits. If there are fewer than $\frac{n
        c R}{\ln 2}$ bits still awaiting transmission, just idle by
        transmitting an arbitrary input letter.

      \item At every noisy channel use, the encoder sends the channel
        input corresponding to the next position in an infinite-length
        random codeword associated with the current data block, where
        the random codewords are drawn iid using the appropriate input
        distribution\footnote{Use the $E_0(\eta)$ maximizing input
        distribution for the $\eta$ such that the data rate $R =
        \frac{E_0(\eta)}{\eta}$.} over the noisy channel's input
        alphabet.

  \item If the time is an integer multiple of $c$, use the noiselessly
        fedback channel outputs to simulate the decoder's attempt to
        decode the current codeword to within a list of the top $2^l$
        items. If the true data-block is one of the $2^l$ items, send
        a $1$ over the noiseless flow-control link. Also send the
        disambiguating $l$ bits representing the true block's index
        within the decoder's list.  Remove the current block of
        $\frac{n c R}{\ln 2}$ bits from the main data queue as
        well. If the true block is not in the decoder's list, just
        send a $0$ over the noiseless flow-control link.

  \item At the decoder, the encoder queue length is known
        perfectly since it can only change by the deterministic
        arrival of data bits or when a noise-free confirm or deny bit
        has been sent over the flow-control link. Thus the decoder
        always knows which input block a given channel output $Y_t$ or
        fortified symbol $S_t$ corresponds to.
  
  \item If the time is an integer multiple of $c$ and the decoder
        receives a $1$ noiselessly, then it decodes what it has seen
        to a list of the top $2^l$ possibilities for this block. It
        will use the next $l$ noisefree flow-control bits to
        disambiguate this list and will use the result as its estimate
        for the block.
\end{itemize}

Such schemes are shown in \cite{OurUpperBoundPaper} to be
asymptotically optimal:
\begin{theorem} \label{thm:fortifiedoptimality}
By appropriate choice of $(n,c,l)$, it is possible to asymptotically
achieve all delay exponents $\alpha < E_{0}(\rho)$ for $R = \frac{E_0(\rho)}{\rho}$ for the fortified system built
around a DMC by adding a rate $\frac{1}{k}$ noisefree
forward flow-control link where $k$ can be made as small as desired.
\end{theorem}
\vspace{0.1in}

\section{Synthesizing a pathway to carry flow-control information}
\label{sec:syntheticflow} 

These codes use time-sharing of the channel to split it into two
parts. One part carries the data and the other part carries flow
control information. 

\subsection{Channels with positive zero-error capacity}
The fortified communication scheme is easily adapted to channels with
strictly positive zero-error capacity by just using the feedback
zero-error capacity to carry the flow-control information
\cite{OurUpperBoundPaper}. There is no $k$. Instead, let $\theta$ be
block-length required to realize feedback zero-error transmission of
at least $l+1$ bits. As illustrated in Figure~\ref{fig:zeroerrorcode},
terminate each chunk with a length $\theta$ feedback zero-error code
and use it to transmit the flow-control information. If the chunk size
is $c$, then it is as though we are operating with only a fraction $(1
- \frac{\theta}{c})$ of the channel uses. The overhead tends to zero
by making the chunk sizes long giving the following corollary to
Theorem~\ref{thm:fortifiedoptimality}:

\begin{figure}
\begin{center}
\setlength{\unitlength}{4000sp}%
\begingroup\makeatletter\ifx\SetFigFont\undefined%
\gdef\SetFigFont#1#2#3#4#5{%
  \reset@font\fontsize{#1}{#2pt}%
  \fontfamily{#3}\fontseries{#4}\fontshape{#5}%
  \selectfont}%
\fi\endgroup%
\begin{picture}(7674,2284)(-11,-1760)
\thinlines
{\color[rgb]{0,0,0}\put(1651,-136){\line( 1, 0){450}}
}%
{\color[rgb]{0,0,0}\put(1651,-286){\line( 1, 0){450}}
}%
{\color[rgb]{0,0,0}\put(  1,-211){\vector( 1, 0){7650}}
}%
{\color[rgb]{0,0,0}\put(6301,-136){\line( 0,-1){150}}
}%
{\color[rgb]{0,0,0}\multiput(7463,-136)(0.00000,-4.54545){34}{\makebox(1.6667,11.6667){\SetFigFont{5}{6}{\rmdefault}{\mddefault}{\updefault}.}}
}%
{\color[rgb]{0,0,0}\multiput(7426,-136)(0.00000,-4.54545){34}{\makebox(1.6667,11.6667){\SetFigFont{5}{6}{\rmdefault}{\mddefault}{\updefault}.}}
}%
{\color[rgb]{0,0,0}\multiput(7388,-136)(0.00000,-4.54545){34}{\makebox(1.6667,11.6667){\SetFigFont{5}{6}{\rmdefault}{\mddefault}{\updefault}.}}
}%
{\color[rgb]{0,0,0}\multiput(7351,-136)(0.00000,-4.54545){34}{\makebox(1.6667,11.6667){\SetFigFont{5}{6}{\rmdefault}{\mddefault}{\updefault}.}}
}%
{\color[rgb]{0,0,0}\multiput(7313,-136)(0.00000,-4.54545){34}{\makebox(1.6667,11.6667){\SetFigFont{5}{6}{\rmdefault}{\mddefault}{\updefault}.}}
}%
{\color[rgb]{0,0,0}\multiput(7276,-136)(0.00000,-4.54545){34}{\makebox(1.6667,11.6667){\SetFigFont{5}{6}{\rmdefault}{\mddefault}{\updefault}.}}
}%
{\color[rgb]{0,0,0}\multiput(7238,-136)(0.00000,-4.54545){34}{\makebox(1.6667,11.6667){\SetFigFont{5}{6}{\rmdefault}{\mddefault}{\updefault}.}}
}%
{\color[rgb]{0,0,0}\multiput(7201,-136)(0.00000,-4.54545){34}{\makebox(1.6667,11.6667){\SetFigFont{5}{6}{\rmdefault}{\mddefault}{\updefault}.}}
}%
{\color[rgb]{0,0,0}\multiput(7163,-136)(0.00000,-4.54545){34}{\makebox(1.6667,11.6667){\SetFigFont{5}{6}{\rmdefault}{\mddefault}{\updefault}.}}
}%
{\color[rgb]{0,0,0}\multiput(7126,-136)(0.00000,-4.54545){34}{\makebox(1.6667,11.6667){\SetFigFont{5}{6}{\rmdefault}{\mddefault}{\updefault}.}}
}%
{\color[rgb]{0,0,0}\multiput(7088,-136)(0.00000,-4.54545){34}{\makebox(1.6667,11.6667){\SetFigFont{5}{6}{\rmdefault}{\mddefault}{\updefault}.}}
}%
{\color[rgb]{0,0,0}\multiput(7051,-136)(0.00000,-4.54545){34}{\makebox(1.6667,11.6667){\SetFigFont{5}{6}{\rmdefault}{\mddefault}{\updefault}.}}
}%
{\color[rgb]{0,0,0}\multiput(7013,-136)(0.00000,-4.54545){34}{\makebox(1.6667,11.6667){\SetFigFont{5}{6}{\rmdefault}{\mddefault}{\updefault}.}}
}%
{\color[rgb]{0,0,0}\multiput(6976,-136)(0.00000,-4.54545){34}{\makebox(1.6667,11.6667){\SetFigFont{5}{6}{\rmdefault}{\mddefault}{\updefault}.}}
}%
{\color[rgb]{0,0,0}\multiput(6938,-136)(0.00000,-4.54545){34}{\makebox(1.6667,11.6667){\SetFigFont{5}{6}{\rmdefault}{\mddefault}{\updefault}.}}
}%
{\color[rgb]{0,0,0}\multiput(6901,-136)(0.00000,-4.54545){34}{\makebox(1.6667,11.6667){\SetFigFont{5}{6}{\rmdefault}{\mddefault}{\updefault}.}}
}%
{\color[rgb]{0,0,0}\multiput(6863,-136)(0.00000,-4.54545){34}{\makebox(1.6667,11.6667){\SetFigFont{5}{6}{\rmdefault}{\mddefault}{\updefault}.}}
}%
{\color[rgb]{0,0,0}\multiput(6826,-136)(0.00000,-4.54545){34}{\makebox(1.6667,11.6667){\SetFigFont{5}{6}{\rmdefault}{\mddefault}{\updefault}.}}
}%
{\color[rgb]{0,0,0}\multiput(6788,-136)(0.00000,-4.54545){34}{\makebox(1.6667,11.6667){\SetFigFont{5}{6}{\rmdefault}{\mddefault}{\updefault}.}}
}%
{\color[rgb]{0,0,0}\multiput(6751,-136)(0.00000,-4.54545){34}{\makebox(1.6667,11.6667){\SetFigFont{5}{6}{\rmdefault}{\mddefault}{\updefault}.}}
}%
{\color[rgb]{0,0,0}\multiput(6713,-136)(0.00000,-4.54545){34}{\makebox(1.6667,11.6667){\SetFigFont{5}{6}{\rmdefault}{\mddefault}{\updefault}.}}
}%
{\color[rgb]{0,0,0}\multiput(6676,-136)(0.00000,-4.54545){34}{\makebox(1.6667,11.6667){\SetFigFont{5}{6}{\rmdefault}{\mddefault}{\updefault}.}}
}%
{\color[rgb]{0,0,0}\multiput(6638,-136)(0.00000,-4.54545){34}{\makebox(1.6667,11.6667){\SetFigFont{5}{6}{\rmdefault}{\mddefault}{\updefault}.}}
}%
{\color[rgb]{0,0,0}\multiput(6601,-136)(0.00000,-4.54545){34}{\makebox(1.6667,11.6667){\SetFigFont{5}{6}{\rmdefault}{\mddefault}{\updefault}.}}
}%
{\color[rgb]{0,0,0}\multiput(6563,-136)(0.00000,-4.54545){34}{\makebox(1.6667,11.6667){\SetFigFont{5}{6}{\rmdefault}{\mddefault}{\updefault}.}}
}%
{\color[rgb]{0,0,0}\multiput(6526,-136)(0.00000,-4.54545){34}{\makebox(1.6667,11.6667){\SetFigFont{5}{6}{\rmdefault}{\mddefault}{\updefault}.}}
}%
{\color[rgb]{0,0,0}\multiput(6488,-136)(0.00000,-4.54545){34}{\makebox(1.6667,11.6667){\SetFigFont{5}{6}{\rmdefault}{\mddefault}{\updefault}.}}
}%
{\color[rgb]{0,0,0}\multiput(6451,-136)(0.00000,-4.54545){34}{\makebox(1.6667,11.6667){\SetFigFont{5}{6}{\rmdefault}{\mddefault}{\updefault}.}}
}%
{\color[rgb]{0,0,0}\multiput(6413,-136)(0.00000,-4.54545){34}{\makebox(1.6667,11.6667){\SetFigFont{5}{6}{\rmdefault}{\mddefault}{\updefault}.}}
}%
{\color[rgb]{0,0,0}\multiput(6376,-136)(0.00000,-4.54545){34}{\makebox(1.6667,11.6667){\SetFigFont{5}{6}{\rmdefault}{\mddefault}{\updefault}.}}
}%
{\color[rgb]{0,0,0}\multiput(6338,-136)(0.00000,-4.54545){34}{\makebox(1.6667,11.6667){\SetFigFont{5}{6}{\rmdefault}{\mddefault}{\updefault}.}}
}%
{\color[rgb]{0,0,0}\put(451,-136){\line( 0,-1){150}}
}%
{\color[rgb]{0,0,0}\put(526,-136){\line( 0,-1){150}}
}%
{\color[rgb]{0,0,0}\put(301,-136){\line( 0,-1){150}}
}%
{\color[rgb]{0,0,0}\put(376,-136){\line( 0,-1){150}}
}%
{\color[rgb]{0,0,0}\put(  1,-136){\line( 0,-1){150}}
}%
{\color[rgb]{0,0,0}\put( 76,-136){\line( 0,-1){150}}
}%
{\color[rgb]{0,0,0}\put(151,-136){\line( 0,-1){150}}
}%
{\color[rgb]{0,0,0}\put(226,-136){\line( 0,-1){150}}
}%
{\color[rgb]{0,0,0}\put(188,-136){\line( 0,-1){150}}
}%
{\color[rgb]{0,0,0}\put(263,-136){\line( 0,-1){150}}
}%
{\color[rgb]{0,0,0}\put( 38,-136){\line( 0,-1){150}}
}%
{\color[rgb]{0,0,0}\put(113,-136){\line( 0,-1){150}}
}%
{\color[rgb]{0,0,0}\put(338,-136){\line( 0,-1){150}}
}%
{\color[rgb]{0,0,0}\put(413,-136){\line( 0,-1){150}}
}%
{\color[rgb]{0,0,0}\put(488,-136){\line( 0,-1){150}}
}%
{\color[rgb]{0,0,0}\put(563,-136){\line( 0,-1){150}}
}%
{\color[rgb]{0,0,0}\put(601,-136){\line( 0,-1){150}}
}%
{\color[rgb]{0,0,0}\put(676,-136){\line( 0,-1){150}}
}%
{\color[rgb]{0,0,0}\put(638,-136){\line( 0,-1){150}}
}%
{\color[rgb]{0,0,0}\put(713,-136){\line( 0,-1){150}}
}%
{\color[rgb]{0,0,0}\put(751,-136){\line( 0,-1){150}}
}%
{\color[rgb]{0,0,0}\put(826,-136){\line( 0,-1){150}}
}%
{\color[rgb]{0,0,0}\put(788,-136){\line( 0,-1){150}}
}%
{\color[rgb]{0,0,0}\put(863,-136){\line( 0,-1){150}}
}%
{\color[rgb]{0,0,0}\put(901,-136){\line( 0,-1){150}}
}%
{\color[rgb]{0,0,0}\put(976,-136){\line( 0,-1){150}}
}%
{\color[rgb]{0,0,0}\put(938,-136){\line( 0,-1){150}}
}%
{\color[rgb]{0,0,0}\put(1013,-136){\line( 0,-1){150}}
}%
{\color[rgb]{0,0,0}\put(1088,-136){\line( 0,-1){150}}
}%
{\color[rgb]{0,0,0}\put(1163,-136){\line( 0,-1){150}}
}%
{\color[rgb]{0,0,0}\put(1051,-136){\line( 0,-1){150}}
}%
{\color[rgb]{0,0,0}\put(1126,-136){\line( 0,-1){150}}
}%
{\color[rgb]{0,0,0}\put(1238,-136){\line( 0,-1){150}}
}%
{\color[rgb]{0,0,0}\put(1313,-136){\line( 0,-1){150}}
}%
{\color[rgb]{0,0,0}\put(1388,-136){\line( 0,-1){150}}
}%
{\color[rgb]{0,0,0}\put(1463,-136){\line( 0,-1){150}}
}%
{\color[rgb]{0,0,0}\put(1201,-136){\line( 0,-1){150}}
}%
{\color[rgb]{0,0,0}\put(1276,-136){\line( 0,-1){150}}
}%
{\color[rgb]{0,0,0}\put(1351,-136){\line( 0,-1){150}}
}%
{\color[rgb]{0,0,0}\put(1426,-136){\line( 0,-1){150}}
}%
{\color[rgb]{0,0,0}\put(1501,-136){\line( 0,-1){150}}
}%
{\color[rgb]{0,0,0}\put(1576,-136){\line( 0,-1){150}}
}%
{\color[rgb]{0,0,0}\put(1651,-136){\line( 0,-1){150}}
}%
{\color[rgb]{0,0,0}\put(1726,-136){\line( 0,-1){150}}
}%
{\color[rgb]{0,0,0}\put(1688,-136){\line( 0,-1){150}}
}%
{\color[rgb]{0,0,0}\put(1763,-136){\line( 0,-1){150}}
}%
{\color[rgb]{0,0,0}\put(1538,-136){\line( 0,-1){150}}
}%
{\color[rgb]{0,0,0}\put(1613,-136){\line( 0,-1){150}}
}%
{\color[rgb]{0,0,0}\put(1838,-136){\line( 0,-1){150}}
}%
{\color[rgb]{0,0,0}\put(1913,-136){\line( 0,-1){150}}
}%
{\color[rgb]{0,0,0}\put(1988,-136){\line( 0,-1){150}}
}%
{\color[rgb]{0,0,0}\put(2063,-136){\line( 0,-1){150}}
}%
{\color[rgb]{0,0,0}\put(1801,-136){\line( 0,-1){150}}
}%
{\color[rgb]{0,0,0}\put(1876,-136){\line( 0,-1){150}}
}%
{\color[rgb]{0,0,0}\put(1951,-136){\line( 0,-1){150}}
}%
{\color[rgb]{0,0,0}\put(2026,-136){\line( 0,-1){150}}
}%
{\color[rgb]{0,0,0}\put(2101,-136){\line( 0,-1){150}}
}%
{\color[rgb]{0,0,0}\put(2176,-136){\line( 0,-1){150}}
}%
{\color[rgb]{0,0,0}\put(2251,-136){\line( 0,-1){150}}
}%
{\color[rgb]{0,0,0}\put(2326,-136){\line( 0,-1){150}}
}%
{\color[rgb]{0,0,0}\put(2138,-136){\line( 0,-1){150}}
}%
{\color[rgb]{0,0,0}\put(2213,-136){\line( 0,-1){150}}
}%
{\color[rgb]{0,0,0}\put(2288,-136){\line( 0,-1){150}}
}%
{\color[rgb]{0,0,0}\put(2363,-136){\line( 0,-1){150}}
}%
{\color[rgb]{0,0,0}\put(2438,-136){\line( 0,-1){150}}
}%
{\color[rgb]{0,0,0}\put(2513,-136){\line( 0,-1){150}}
}%
{\color[rgb]{0,0,0}\put(2588,-136){\line( 0,-1){150}}
}%
{\color[rgb]{0,0,0}\put(2663,-136){\line( 0,-1){150}}
}%
{\color[rgb]{0,0,0}\put(2738,-136){\line( 0,-1){150}}
}%
{\color[rgb]{0,0,0}\put(2813,-136){\line( 0,-1){150}}
}%
{\color[rgb]{0,0,0}\put(2888,-136){\line( 0,-1){150}}
}%
{\color[rgb]{0,0,0}\put(2963,-136){\line( 0,-1){150}}
}%
{\color[rgb]{0,0,0}\put(2401,-136){\line( 0,-1){150}}
}%
{\color[rgb]{0,0,0}\put(2476,-136){\line( 0,-1){150}}
}%
{\color[rgb]{0,0,0}\put(2551,-136){\line( 0,-1){150}}
}%
{\color[rgb]{0,0,0}\put(2626,-136){\line( 0,-1){150}}
}%
{\color[rgb]{0,0,0}\put(2701,-136){\line( 0,-1){150}}
}%
{\color[rgb]{0,0,0}\put(2776,-136){\line( 0,-1){150}}
}%
{\color[rgb]{0,0,0}\put(2851,-136){\line( 0,-1){150}}
}%
{\color[rgb]{0,0,0}\put(2926,-136){\line( 0,-1){150}}
}%
{\color[rgb]{0,0,0}\put(3151,-136){\line( 0,-1){150}}
}%
{\color[rgb]{0,0,0}\put(3226,-136){\line( 0,-1){150}}
}%
{\color[rgb]{0,0,0}\put(3038,-136){\line( 0,-1){150}}
}%
{\color[rgb]{0,0,0}\put(3113,-136){\line( 0,-1){150}}
}%
{\color[rgb]{0,0,0}\put(3188,-136){\line( 0,-1){150}}
}%
{\color[rgb]{0,0,0}\put(3263,-136){\line( 0,-1){150}}
}%
{\color[rgb]{0,0,0}\put(3301,-136){\line( 0,-1){150}}
}%
{\color[rgb]{0,0,0}\put(3376,-136){\line( 0,-1){150}}
}%
{\color[rgb]{0,0,0}\put(3001,-136){\line( 0,-1){150}}
}%
{\color[rgb]{0,0,0}\put(3076,-136){\line( 0,-1){150}}
}%
{\color[rgb]{0,0,0}\put(3338,-136){\line( 0,-1){150}}
}%
{\color[rgb]{0,0,0}\put(3413,-136){\line( 0,-1){150}}
}%
{\color[rgb]{0,0,0}\put(3451,-136){\line( 0,-1){150}}
}%
{\color[rgb]{0,0,0}\put(3526,-136){\line( 0,-1){150}}
}%
{\color[rgb]{0,0,0}\put(3488,-136){\line( 0,-1){150}}
}%
{\color[rgb]{0,0,0}\put(3563,-136){\line( 0,-1){150}}
}%
{\color[rgb]{0,0,0}\put(3751,-136){\line( 0,-1){150}}
}%
{\color[rgb]{0,0,0}\put(3826,-136){\line( 0,-1){150}}
}%
{\color[rgb]{0,0,0}\put(3601,-136){\line( 0,-1){150}}
}%
{\color[rgb]{0,0,0}\put(3676,-136){\line( 0,-1){150}}
}%
{\color[rgb]{0,0,0}\put(4088,-136){\line( 0,-1){150}}
}%
{\color[rgb]{0,0,0}\put(4163,-136){\line( 0,-1){150}}
}%
{\color[rgb]{0,0,0}\put(4238,-136){\line( 0,-1){150}}
}%
{\color[rgb]{0,0,0}\put(4313,-136){\line( 0,-1){150}}
}%
{\color[rgb]{0,0,0}\put(4201,-136){\line( 0,-1){150}}
}%
{\color[rgb]{0,0,0}\put(4276,-136){\line( 0,-1){150}}
}%
{\color[rgb]{0,0,0}\put(3938,-136){\line( 0,-1){150}}
}%
{\color[rgb]{0,0,0}\put(4013,-136){\line( 0,-1){150}}
}%
{\color[rgb]{0,0,0}\put(3638,-136){\line( 0,-1){150}}
}%
{\color[rgb]{0,0,0}\put(3713,-136){\line( 0,-1){150}}
}%
{\color[rgb]{0,0,0}\put(3788,-136){\line( 0,-1){150}}
}%
{\color[rgb]{0,0,0}\put(3863,-136){\line( 0,-1){150}}
}%
{\color[rgb]{0,0,0}\put(3901,-136){\line( 0,-1){150}}
}%
{\color[rgb]{0,0,0}\put(3976,-136){\line( 0,-1){150}}
}%
{\color[rgb]{0,0,0}\put(4051,-136){\line( 0,-1){150}}
}%
{\color[rgb]{0,0,0}\put(4126,-136){\line( 0,-1){150}}
}%
{\color[rgb]{0,0,0}\put(4351,-136){\line( 0,-1){150}}
}%
{\color[rgb]{0,0,0}\put(4426,-136){\line( 0,-1){150}}
}%
{\color[rgb]{0,0,0}\put(4388,-136){\line( 0,-1){150}}
}%
{\color[rgb]{0,0,0}\put(4463,-136){\line( 0,-1){150}}
}%
{\color[rgb]{0,0,0}\put(4501,-136){\line( 0,-1){150}}
}%
{\color[rgb]{0,0,0}\put(4576,-136){\line( 0,-1){150}}
}%
{\color[rgb]{0,0,0}\put(4651,-136){\line( 0,-1){150}}
}%
{\color[rgb]{0,0,0}\put(4726,-136){\line( 0,-1){150}}
}%
{\color[rgb]{0,0,0}\put(4538,-136){\line( 0,-1){150}}
}%
{\color[rgb]{0,0,0}\put(4613,-136){\line( 0,-1){150}}
}%
{\color[rgb]{0,0,0}\put(4688,-136){\line( 0,-1){150}}
}%
{\color[rgb]{0,0,0}\put(4763,-136){\line( 0,-1){150}}
}%
{\color[rgb]{0,0,0}\put(5138,-136){\line( 0,-1){150}}
}%
{\color[rgb]{0,0,0}\put(5213,-136){\line( 0,-1){150}}
}%
{\color[rgb]{0,0,0}\put(4988,-136){\line( 0,-1){150}}
}%
{\color[rgb]{0,0,0}\put(5063,-136){\line( 0,-1){150}}
}%
{\color[rgb]{0,0,0}\put(4838,-136){\line( 0,-1){150}}
}%
{\color[rgb]{0,0,0}\put(4913,-136){\line( 0,-1){150}}
}%
{\color[rgb]{0,0,0}\put(4951,-136){\line( 0,-1){150}}
}%
{\color[rgb]{0,0,0}\put(5026,-136){\line( 0,-1){150}}
}%
{\color[rgb]{0,0,0}\put(4801,-136){\line( 0,-1){150}}
}%
{\color[rgb]{0,0,0}\put(4876,-136){\line( 0,-1){150}}
}%
{\color[rgb]{0,0,0}\put(5101,-136){\line( 0,-1){150}}
}%
{\color[rgb]{0,0,0}\put(5176,-136){\line( 0,-1){150}}
}%
{\color[rgb]{0,0,0}\put(5251,-136){\line( 0,-1){150}}
}%
{\color[rgb]{0,0,0}\put(5326,-136){\line( 0,-1){150}}
}%
{\color[rgb]{0,0,0}\put(5288,-136){\line( 0,-1){150}}
}%
{\color[rgb]{0,0,0}\put(5363,-136){\line( 0,-1){150}}
}%
{\color[rgb]{0,0,0}\put(5438,-136){\line( 0,-1){150}}
}%
{\color[rgb]{0,0,0}\put(5513,-136){\line( 0,-1){150}}
}%
{\color[rgb]{0,0,0}\put(5551,-136){\line( 0,-1){150}}
}%
{\color[rgb]{0,0,0}\put(5626,-136){\line( 0,-1){150}}
}%
{\color[rgb]{0,0,0}\put(5588,-136){\line( 0,-1){150}}
}%
{\color[rgb]{0,0,0}\put(5663,-136){\line( 0,-1){150}}
}%
{\color[rgb]{0,0,0}\put(5401,-136){\line( 0,-1){150}}
}%
{\color[rgb]{0,0,0}\put(5476,-136){\line( 0,-1){150}}
}%
{\color[rgb]{0,0,0}\put(5701,-136){\line( 0,-1){150}}
}%
{\color[rgb]{0,0,0}\put(5776,-136){\line( 0,-1){150}}
}%
{\color[rgb]{0,0,0}\put(6151,-136){\line( 0,-1){150}}
}%
{\color[rgb]{0,0,0}\put(6226,-136){\line( 0,-1){150}}
}%
{\color[rgb]{0,0,0}\put(6001,-136){\line( 0,-1){150}}
}%
{\color[rgb]{0,0,0}\put(6076,-136){\line( 0,-1){150}}
}%
{\color[rgb]{0,0,0}\put(5851,-136){\line( 0,-1){150}}
}%
{\color[rgb]{0,0,0}\put(5926,-136){\line( 0,-1){150}}
}%
{\color[rgb]{0,0,0}\put(5738,-136){\line( 0,-1){150}}
}%
{\color[rgb]{0,0,0}\put(5813,-136){\line( 0,-1){150}}
}%
{\color[rgb]{0,0,0}\put(6188,-136){\line( 0,-1){150}}
}%
{\color[rgb]{0,0,0}\put(6263,-136){\line( 0,-1){150}}
}%
{\color[rgb]{0,0,0}\put(6038,-136){\line( 0,-1){150}}
}%
{\color[rgb]{0,0,0}\put(6113,-136){\line( 0,-1){150}}
}%
{\color[rgb]{0,0,0}\put(5888,-136){\line( 0,-1){150}}
}%
{\color[rgb]{0,0,0}\put(5963,-136){\line( 0,-1){150}}
}%
{\color[rgb]{0,0,0}\put(4201,-136){\line(-1, 0){450}}
}%
{\color[rgb]{0,0,0}\put(4201,-286){\line(-1, 0){450}}
}%
{\color[rgb]{0,0,0}\put(5851,-136){\line( 1, 0){450}}
}%
{\color[rgb]{0,0,0}\put(5851,-286){\line( 1, 0){450}}
}%
{\color[rgb]{0,0,0}\put(1651,-136){\line( 3,-1){450}}
}%
{\color[rgb]{0,0,0}\put(1651,-286){\line( 3, 1){450}}
}%
{\color[rgb]{0,0,0}\put(3751,-286){\line( 3, 1){450}}
}%
{\color[rgb]{0,0,0}\put(3751,-136){\line( 3,-1){450}}
}%
{\color[rgb]{0,0,0}\put(5851,-136){\line( 3,-1){450}}
}%
{\color[rgb]{0,0,0}\put(5851,-286){\line( 3, 1){450}}
}%
{\color[rgb]{0,0,0}\multiput(1651,-286)(0.00000,-4.50000){151}{\makebox(1.6667,11.6667){\SetFigFont{5}{6}{\rmdefault}{\mddefault}{\updefault}.}}
}%
{\color[rgb]{0,0,0}\multiput(2101,-286)(0.00000,-4.50000){151}{\makebox(1.6667,11.6667){\SetFigFont{5}{6}{\rmdefault}{\mddefault}{\updefault}.}}
}%
{\color[rgb]{0,0,0}\multiput(3751,-286)(0.00000,-4.50000){151}{\makebox(1.6667,11.6667){\SetFigFont{5}{6}{\rmdefault}{\mddefault}{\updefault}.}}
}%
{\color[rgb]{0,0,0}\multiput(4201,-286)(0.00000,-4.50000){151}{\makebox(1.6667,11.6667){\SetFigFont{5}{6}{\rmdefault}{\mddefault}{\updefault}.}}
}%
{\color[rgb]{0,0,0}\multiput(5851,-286)(0.00000,-4.50000){151}{\makebox(1.6667,11.6667){\SetFigFont{5}{6}{\rmdefault}{\mddefault}{\updefault}.}}
}%
{\color[rgb]{0,0,0}\multiput(6301,-286)(0.00000,-4.50000){151}{\makebox(1.6667,11.6667){\SetFigFont{5}{6}{\rmdefault}{\mddefault}{\updefault}.}}
}%
{\color[rgb]{0,0,0}\put(3976,-1561){\vector( 0, 1){675}}
}%
{\color[rgb]{0,0,0}\put(3976,-1486){\line( 1, 0){2100}}
\put(6076,-1486){\vector( 0, 1){225}}
}%
{\color[rgb]{0,0,0}\put(3976,-1486){\line(-1, 0){2100}}
\put(1876,-1486){\vector( 0, 1){600}}
}%
{\color[rgb]{0,0,0}\put(  1,-511){\line( 0,-1){150}}
}%
{\color[rgb]{0,0,0}\put(2101,-511){\line( 0,-1){150}}
}%
{\color[rgb]{0,0,0}\put(2101,-586){\line( 1, 0){2100}}
}%
{\color[rgb]{0,0,0}\put(2101,-511){\line( 0,-1){150}}
}%
{\color[rgb]{0,0,0}\put(4201,-511){\line( 0,-1){150}}
}%
{\color[rgb]{0,0,0}\put(4201,-586){\line( 1, 0){2100}}
}%
{\color[rgb]{0,0,0}\put(4201,-511){\line( 0,-1){150}}
}%
{\color[rgb]{0,0,0}\put(6301,-511){\line( 0,-1){150}}
}%
{\color[rgb]{0,0,0}\put(  1,-586){\line( 1, 0){2100}}
}%
\put(6076,-1186){\makebox(0,0)[b]{\smash{\SetFigFont{7}{6}{\rmdefault}{\mddefault}{\updefault}{\color[rgb]{0,0,0}disambiguation}%
}}}
\put(3976,-1711){\makebox(0,0)[b]{\smash{\SetFigFont{7}{6}{\rmdefault}{\mddefault}{\updefault}{\color[rgb]{0,0,0}$(l+1)$-bit zero-error feedback  block codes}%
}}}
\put(  1,389){\makebox(0,0)[lb]{\smash{\SetFigFont{7}{6}{\rmdefault}{\mddefault}{\updefault}{\color[rgb]{0,0,0}DMC channel uses}%
}}}
\put(1876, 14){\makebox(0,0)[b]{\smash{\SetFigFont{7}{6}{\rmdefault}{\mddefault}{\updefault}{\color[rgb]{0,0,0}$\theta$}%
}}}
\put(3976, 14){\makebox(0,0)[b]{\smash{\SetFigFont{7}{6}{\rmdefault}{\mddefault}{\updefault}{\color[rgb]{0,0,0}$\theta$}%
}}}
\put(6076, 14){\makebox(0,0)[b]{\smash{\SetFigFont{7}{6}{\rmdefault}{\mddefault}{\updefault}{\color[rgb]{0,0,0}$\theta$}%
}}}
\put(1876,-811){\makebox(0,0)[b]{\smash{\SetFigFont{7}{6}{\rmdefault}{\mddefault}{\updefault}{\color[rgb]{0,0,0}deny}%
}}}
\put(3976,-811){\makebox(0,0)[b]{\smash{\SetFigFont{7}{6}{\rmdefault}{\mddefault}{\updefault}{\color[rgb]{0,0,0}deny}%
}}}
\put(6076,-961){\makebox(0,0)[b]{\smash{\SetFigFont{7}{6}{\rmdefault}{\mddefault}{\updefault}{\color[rgb]{0,0,0}confirm +}%
}}}
\put(751, 14){\makebox(0,0)[b]{\smash{\SetFigFont{7}{6}{\rmdefault}{\mddefault}{\updefault}{\color[rgb]{0,0,0}$c-\theta$}%
}}}
\put(2851, 14){\makebox(0,0)[b]{\smash{\SetFigFont{7}{6}{\rmdefault}{\mddefault}{\updefault}{\color[rgb]{0,0,0}$c-\theta$}%
}}}
\put(4951, 14){\makebox(0,0)[b]{\smash{\SetFigFont{7}{6}{\rmdefault}{\mddefault}{\updefault}{\color[rgb]{0,0,0}$c-\theta$}%
}}}
\put(1051,-511){\makebox(0,0)[b]{\smash{\SetFigFont{7}{6}{\rmdefault}{\mddefault}{\updefault}{\color[rgb]{0,0,0}first chunk}%
}}}
\put(3151,-511){\makebox(0,0)[b]{\smash{\SetFigFont{7}{6}{\rmdefault}{\mddefault}{\updefault}{\color[rgb]{0,0,0}second chunk}%
}}}
\put(5251,-511){\makebox(0,0)[b]{\smash{\SetFigFont{7}{6}{\rmdefault}{\mddefault}{\updefault}{\color[rgb]{0,0,0}third chunk}%
}}}
\put(751,-811){\makebox(0,0)[b]{\smash{\SetFigFont{7}{6}{\rmdefault}{\mddefault}{\updefault}{\color[rgb]{0,0,0}random increment}%
}}}
\put(2851,-811){\makebox(0,0)[b]{\smash{\SetFigFont{7}{6}{\rmdefault}{\mddefault}{\updefault}{\color[rgb]{0,0,0}random increment}%
}}}
\put(4951,-811){\makebox(0,0)[b]{\smash{\SetFigFont{7}{6}{\rmdefault}{\mddefault}{\updefault}{\color[rgb]{0,0,0}random increment}%
}}}
\end{picture}
\end{center}
\caption{One block's transmission in the $(n,c,l,\theta)$ code for
  noisy channels. The $\theta$ is used to carry $l+1$ flow-control
  bits reliably. If the channel has a strictly positive feedback
  zero-error capacity, $\theta$ does not scale with $c$. If it does
  not, $\theta$ is proportional to $c$.}
\label{fig:zeroerrorcode}
\end{figure}

\begin{corollary} \label{cor:zeroerrorcase} By appropriate choice of
  $(n,c,l)$, it is possible to asymptotically achieve all delay
  exponents $\alpha < E_{0}(\rho)$ for $R = \frac{E_0(\rho)}{\rho}$
  for any channel with $C_{0,f} > 0$. 
\end{corollary}

\subsection{Channels without zero-error capacity}

When the channel has no zero error capacity, then we can still
allocate $\theta$ channel uses per chunk to carry flow control
information and have the encoder just assume that it was received
correctly. This can be done by using an infinite constraint-length
time-varying random convolutional code.\footnote{See
  \cite{OurUpperBoundPaper} to see some tricks that allow such a code
  to be operated with feedback at bounded expected computational
  complexity, at least at low rates.} This gives a delay-universal
scheme that is guaranteed to eventually get the flow-control
information across correctly. Unlike a zero-error code, all that such
a code can guarantee is that the probability of error in the entire
message stream prefix is exponentially small in the number of channel
uses that have occurred in the code since that message stream prefix
was determined.

The flow-control information can be viewed as low-rate ``punctuation''
that tells the decoder how to parse the channel outputs that are
carrying the data itself. Essentially, the punctuation gives
``commas'' that separate out the different message blocks\footnote{As
  well as disambiguating any lists.}. Here, we assume that the decoder
uses its current best estimate of the punctuation to re-parse the
history of the data-carrying stream. Then the data-carrying channel
outputs are decoded assuming that the flow-control information is
correct. Any bits that have reached their deadlines are emitted, but
this does not prevent the decoder from re-parsing them in the future. 

Consequently, an error can occur at the decoder in two different ways.
As before, the data-carrying stream could be corrupted due to channel
atypicality in those slots. However, the flow-control stream could
also become corrupted. As a result, the $\theta$ must be kept
proportional to the chunk length $c$ to avoid having the flow-control
messages cause too many errors. The effective rate of the flow control
information therefore goes to zero as $c \rightarrow \infty$ and the
relevant error exponent is about $E_0(1)$.  Balancing the error
probabilities and optimizing over the choice of $\theta$ gives the
following theorem: 

\begin{theorem} \label{thm:genericachieve}
By appropriate choice of $(n,c,l,\theta)$, it is possible to
asymptotically achieve all delay exponents $\alpha < E'(R)$ where the
tradeoff curve is given parametrically by varying $\rho \in (0,\infty)$:
\begin{eqnarray} 
E'(\rho) & = & \left(\frac{1}{E_0(\rho)} + \frac{1}{E_0(1)}\right)^{-1} \\
R(\rho)  & = & \frac{E'(\rho)}{\rho}
\end{eqnarray}
\end{theorem} {\em Proof:} For simplicity of exposition, we assume
that the block length $n$, chunk size $c$, and list size $l$ are large
enough that the code essentially achieves the focusing bound for
whatever the effective rate is. The various $\epsilon$ terms are
ignored.

Let $\psi$ be the proportion of channel uses dedicated as overhead to
run the low-rate flow-control channel. So the effective chunk size in
the data-stream is $c' = c (1-\psi)$. The effective rate of the
message stream is thereby increased to $\frac{R}{1-\psi}$. Assuming
that the flow control information is correct, the delay-universal
error exponent is thus $E_a(\frac{R}{1-\psi})$ with respect to the
delay in terms of code channel uses. But there are only $(1-\psi)$
code channel uses per unit of actual time and so the delay exponent is
$(1-\psi)E_a(\frac{R}{1-\psi})$ with respect to true delay.

Meanwhile $\theta = c \psi$. The effective flow-control information
rate is $\frac{l+1}{c \psi} \approx 0$ since $c$ can be made as large
as we want. Since this code achieves the random-coding error exponent,
the delay-universal error exponent for the flow-control stream is 
essentially $E_0(1)$ with flow-code-channel uses since that is the
zero-rate point for random coding. But there are only $\psi$
flow-code-channel uses per actual time and so the delay-exponent for the
flow-control stream is actually $\psi E_0(1)$ with respect to true
delay.

Pick a fixed-delay $d$ large. It can be written as $d = d_f + d_m$ in
$d$ different ways. Let $d_f$ be the part of the delay that is
``burned'' by the flow-control stream. Thus, with probability
exponentially small in $d_f$, this suffix of time has possibly
incorrect flow-control information and so can not be trusted to be
interpreted correctly. Thus, the performance of the code with delay
$d$ is like the performance of the underlying $(n,c',l)$ code with
delay $d_m$. Since the channel uses are disjoint, the two error events
are independent and thus the achieved exponent is the weighted average
of the two error exponents. Balancing the exponents of the two parts
tells us to set:
$$E' = \psi E_0(1) = (1-\psi)E_a(\frac{R}{1-\psi})$$
with the resulting error probability with delay being governed by
$\approx d \exp(-d \psi E_0(1))$. The polynomial term $d$ in front is
dominated entirely by the exponential decay and can be ignored.

Using the parametric forms using $\rho$ for $E_a$, we get a pair of
equations:
\begin{eqnarray}
\psi E_0(1) & = & (1-\psi)E_0(\rho) \\
\frac{E_0(\rho)}{\rho} & = & \frac{R}{1-\psi} 
\end{eqnarray}
The first thing to notice is that simple substitution gives
$$R = \frac{(1-\psi)E_0(\rho)}{\rho} = \frac{\psi E_0(1)}{\rho} = \frac{E'}{\rho}$$

Solving for $\psi$ shows (after a little algebra) that 
\begin{equation} \label{eqn:psidef}
\psi = \frac{E_0(\rho)}{E_0(1) + E_0(\rho)}
\end{equation}
This way $1-\psi = \frac{E_0(1)}{E_0(1) + E_0(\rho)}$ and the first
equation is clearly true. Similarly $\frac{1}{1 - \psi} = 1 +
\frac{E_0(\rho)}{E_0(1)}$ and $\frac{\psi}{1 - \psi} =
\frac{E_0(\rho)}{E_0(1)}$ and thus the second equation is also true.  

\begin{eqnarray*}
E' & = & \psi E_0(1) \\
& = & \frac{E_0(\rho) E_0(1)}{E_0(1) + E_0(\rho)} \\
& = & \left(\frac{1}{E_0(\rho)} + \frac{1}{E_0(1)}\right)^{-1}
\end{eqnarray*}
Which establishes the theorem. \hfill $\Box$

The superiority of these exponents to the sphere-packing bound in the
high rate regime is immediately clear since they are basically like
the focusing bound in form. Some algebra and simple calculus reveals
that the focusing bound\footnote{When $\frac{\partial^2
  E_0(0)}{\partial \rho^2}=0$, page 143 in \cite{gallager} reveals
that the Sphere-packing bound is a straight line hitting zero at
capacity. In such cases, the focusing-bound is bounded away from zero
even in the neighborhood of capacity and hence this curve has an
infinite slope.} has slope $2C/ \frac{\partial^2
  E_0(0)}{\partial \rho^2}$ in the vicinity of the $(C,0)$ point,
while the $E'(R)$ curve achieved by Theorem~\ref{thm:genericachieve}
has the lower slope $E_0(1) / (C - \frac{E_0(1)}{2C}
\left(\frac{\partial^2 E_0(0)}{\partial \rho^2} \right))$. Either way,
in generic cases, the reliability drops linearly in the neighborhood of
capacity rather than in a quadratically flat manner. 

\begin{figure}
\begin{center}
\includegraphics[width=5.25in,height=3.75in]{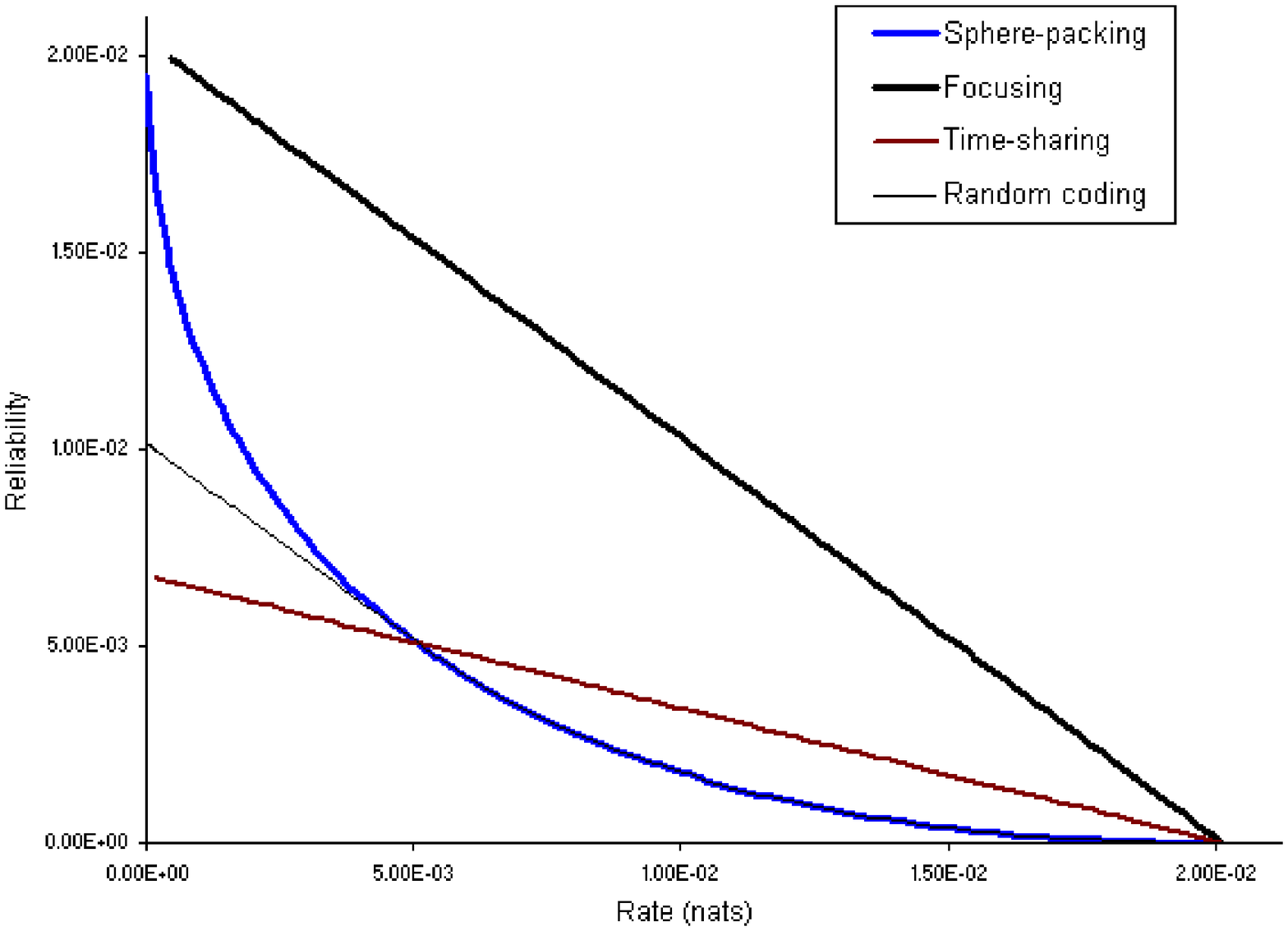}
\end{center}
\caption{The reliability functions for the binary symmetric channel
with crossover probability $\delta=0.4$. The sphere-packing bound
approaches capacity quadratically flat while the focusing bound and
the new scheme both approach the capacity point linearly.}
\label{fig:beatspherebsc}
\end{figure}

Figure~\ref{fig:beatspherebsc} illustrates the bounds for a BSC
with crossover probability $0.4$.

\section{Conclusions}

Even when there is no zero error capacity, flow-control can be used to
substantially beat the sphere-packing bound with respect to delay at
high rates. The arguments from \cite{OurUpperBoundPaper} dealing with
fixed-delay feedback also apply to the new code and show that the
reliabilities achieved here are still asymptotically achievable even
if the feedback is delayed. The key is that our flow-control code does
not need instantaneous feedback to achieve its internal reliability
target $\approx E_0(1)$.

We conjecture that the gap between the focusing bound and the
reliabilities achieved by our scheme in the no-zero-error case is due
to our {\em a-priori} splitting of the channel into dedicated data
and flow-control links. The parallel channel coding advantage tells us
that splitting a channel generally results in a loss of
reliability. The codes in \cite{TuncThesis} performed much better in
the low-rate regime because they had the flow-control information
implicitly within the message stream itself. 

\section*{Acknowledgments}
The author thanks his student Tunc Simsek for many productive
discussions. This work essentially builds on the line of investigation
that we opened up in Tunc's doctoral thesis.

\bibliographystyle{IEEEtran}
\bibliography{IEEEabrv,./MyMainBibliography} 
\end{document}